\documentclass[journal]{new-aiaa}

\usepackage{varioref} 			
\usepackage{wrapfig}			
\usepackage{threeparttable}		
\usepackage{dcolumn}			
\newcolumntype{d}{D{.}{.}{-1}}
\usepackage{nomencl}			
\makeglossary
\usepackage{fancyvrb}			
\fvset{fontsize=\footnotesize,xleftmargin=2em}
\usepackage{color}
\usepackage{xcolor}
\usepackage{lettrine}			
\usepackage[dvips]{dropping} 	
\hypersetup{hidelinks}
\definecolor{Dblue}{RGB}{0,90,158}
\usepackage{psfrag}
\usepackage{soul}
\usepackage{upgreek}
\usepackage{latexsym}
\usepackage{amsmath}
\usepackage{bm}
\usepackage{epsfig}
\usepackage{graphicx}
\usepackage[export]{adjustbox}
\usepackage{arydshln}
\usepackage{array}
\usepackage{multirow} 
\usepackage{xfrac}
\usepackage{tikz}
\usetikzlibrary{arrows.meta}
\usetikzlibrary{decorations}
\usetikzlibrary{decorations.markings}
\usetikzlibrary{calc,decorations.pathreplacing}
\usepackage{lipsum}
\usepackage{mathtools}
\usepackage{ar}
\usepackage{enumitem}
\usepackage{longtable}
\usepackage{url}
\usepackage{siunitx}
\usepackage{nicefrac}
\usepackage{changepage}
\usepackage[version=4]{mhchem}
\usepackage{longtable,tabularx}
\usetikzlibrary{arrows, arrows.meta,chains,positioning,shapes}
\makeatletter
\tikzset{reset join/.code={\def\tikz@after@path{}}}
\makeatother
 
\title{Low-Frequency Intensity Modulation of High-Frequency Rotor Noise}

\author{Woutijn J. Baars\footnote{Assistant Professor, AIAA Senior Member, \url{w.j.baars@tudelft.nl}, \url{www.baars-delft.tech}} and Daniele Ragni\footnote{Associate Professor, AIAA Member, \url{d.ragni@tudelft.nl} \\[3pt] \indent \indent Presented as AIAA Paper \href{https://doi.org/10.2514/6.2023-3215}{2023-3215} at the AIAA AVIATION 2023 Forum, San Diego, CA, 12-16 June 2023. Copyright \textcopyright~2023 by Baars and Ragni. Submitted to the AIAA Journal.}}
\affil{Faculty of Aerospace Engineering, Delft University of Technology, 2629 HS Delft, The Netherlands}

\begin{document}

\maketitle
\begin{abstract}
Acoustic spectra of rotor noise yield frequency-distributions of energy within pressure time series. However, they are unable to reveal phase-relations between different frequency components, while the latter play a role in low-frequency intensity modulation of higher-frequency rotor noise. Baars \emph{et al.} (AIAA Paper \href{https://doi.org/10.2514/6.2021-0713}{2021-0713}) outlined a methodology to quantify inter-frequency modulation, which in the current work is applied to a comprehensive acoustic dataset of a rotor operating at low Reynolds number at advance ratios ranging from $J = 0$ to $0.61$. The findings strengthen earlier observations for the case of a hovering rotor, in which the modulation of the high-frequency noise is strongest at angles of $\theta \approx -20^\circ$ (below the rotor plane). For the non-zero advance ratios, modulation becomes dominant in the sector $-45^\circ \lesssim \theta \lesssim 0^\circ$, and is maximum in strength for the highest advance ratio tested ($J = 0.61$). Intensity-modulation of high-frequency noise is primarily the consequence of a far-field observer experiencing a cyclic sweep through the noise directivity patterns of the relatively directive trailing-edge/shedding noise component. This noise becomes more intense with increasing J and is associated with the broadband features of the (partially) separated flow over the rotor blades.
\end{abstract}

\section{Introduction and context}\label{sec:intro}
\lettrine{U}{rban} air mobility (UAM) vehicles and drones comprise rotors that are typically smaller than the single-rotor technology of conventional helicopters. For instance, it is not uncommon that the many electrical takeoff and landing (eVTOL) prototype vehicles contain a multitude of rotors, \emph{e.g.}, the Joby Aviation vehicle includes 6 rotors, the EHang 216 autonomous aerial vehicle has 6 rotors, the Supernal SA-1 eVTOL aircraft has 4 tiltrotors and 4 sets of stacked co-rotating rotors, and the VoloDrone and VoloCity vehicles of Volocopter include 18 rotors each. Assessing the rotor noise of new advanced air mobility (AAM) vehicles has gained a high priority, due to their envisioned operation in densely populated areas \cite{tegler:2020a}. At the same time, engineering studies on the noise impact of rotors should be revisited because time-varying aspects of acoustic waveforms are rarely addressed, while these influence the human perception of rotor noise.

Most studies on acoustic aspects of small-scale rotors consider standard characterization schemes that rely on time- and/or ensemble-averaging \cite{ol:2008c,brandt:2011c,sinibaldi:2013a,zawodny:2016c,zawodny:2017c,tinney:2018a,tinney:2019c,alexander:2019c,jordan:2020c,yang:2020a,tinney:2020a,valdez:2022a}: results are typically condensed to a set of acoustic spectra, their integrated energy (overall sound pressure level), as well as the directivity patterns of that acoustic energy. The time-dependent \emph{``wop-wop"} noise component from larger-size rotorcraft, or even the higher-frequency \emph{``buzzing"} noise of drone propellers, is a phenomenon of \emph{time-varying noise intensity} and not necessarily a direct perception of the blade passing frequency (BPF), denoted as $f_b$. Current noise certification standards fall short in capturing time-varying aspects of a noise signal (\emph{e.g.}, the tone-corrected, effective perceived noise level (EPNL) and/or the A-weighted sound exposure level (SEL) in 14\,CFR\,Part 36). Paradoxically, there is a growing body of knowledge that such time-varying aspects are highly relevant for the level of (psycho-acoustic) annoyance \cite{christian:2017c,rizzi:2017c,zawodny:2018c,krishnamurthy:2018c,christian:2019c,yang:2020a,gan:2022c}. Because noise levels that comply with a certification standard are not acceptable to the public per say, it is evident that complementary assessments of noise-perception aspects are needed.

\subsection{Rotor noise modulation}\label{sec:modint}
\noindent Understanding the low-frequency modulation of higher-frequency rotor noise is important for the development of low-order modelling- and auralization-algorithms \cite{krishnamurthy:2019c,han:2020a,li:2020c}, as well as psychoacoustic modelling \cite{torija:2021a,torija:2021ab,torija:2022a}. Before considering the temporal variation of rotor noise, a short review of noise sources is provided. Periodic rotor-noise components are classified as thickness noise and blade loading noise \cite{marte:1970tr,amiet:1977a,glegg:2017bk,candeloro:2019a}. A superimposed component of broadband noise originates from turbulence ingestion (a leading-edge mechanism) and vortical turbulent-boundary layer motions convecting past the trailing edge. In addition, for low Reynolds-number propellers, an additional near-wake source comes from the vortex shedding behind laminar and/or turbulent separation regions \cite{grande:2022a}. And finally, even when the rotor operates in a clean flow, the turbulence-ingestion noise can become dominant through the onset of blade-vortex interaction (BVI). This interaction is determined by the distance between the tip-vortex deployed by consecutive blades and the blades themselves.

\begin{figure*}[htb!] 
\vspace{0pt}
\centering
\includegraphics[width = 0.999\textwidth]{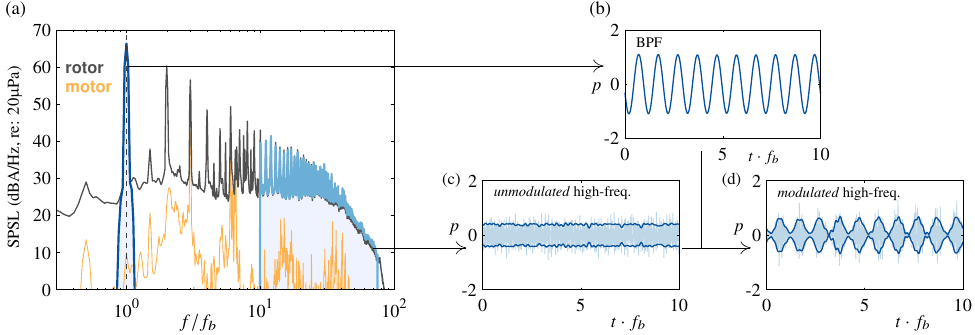}
\caption{(a) Typical acoustic spectrum of rotor noise used to illustrate BPF modulation (BPFM). This spectrum corresponds to a rotor noise time series measured in the current study (microphone B, indicated in Fig.~\ref{fig:expAC}(b), for an advance ratio of $J = 0$). (b,c,d) Time series of the narrow band-pass filtered signal encompassing the BPF, and the high-pass filtered (un)modulated noise residing at $f > 10f_b$.}
\label{fig:modillus}
\end{figure*}

Temporal variations in the acoustic intensity and characteristic frequency of the noise are dubbed intensity (or amplitude) and frequency modulations, respectively. In this work, we consider the intensity modulation and refer to this as \emph{BPF modulation} (BPFM), because it will become evident that its time-scale is prescribed by the rotating motion of the blade. Note that this is fundamentally different from the variations in noise amplitude and Doppler-frequency shift that occur for transient helicopter flyover manoeuvres. In those cases, studies do occasionally employ time-preserving schemes when dealing with non-stationary acoustic signals \cite{stephenson:2014a,rizzi:2019c}; they focus on the acoustic footprints affiliated with \emph{very} long-timescale variations in noise, relative to the BPF, associated with manoeuvres of the flight vehicle. Gan \emph{et al.} \cite{gan:2022c} studied temporal variations of rotor noise from a Bell 206 helicopter in level and descending flight, including effects of aerodynamic interactions and the noise of the tail rotor. Here we study the effect in a more fundamental setting, using an isolated rotor. To illustrate BPFM, consider an acoustic spectrum of rotor noise shown in Fig.~\ref{fig:modillus}(a). The acoustic pressure time series associated with the BPF can be generated using a narrow band-pass filter and is shown in Fig.~\ref{fig:modillus}(b). When the high-frequency content of the signal is \emph{unmodulated}, the time series after high-pass filtering (in this example $f > 10f_b$) has a time-invariant envelope of the intensity (Fig.~\ref{fig:modillus}c). However, due to the rotating nature of the blade's noise sources, a \emph{modulated} intensity-envelope may arise. This modulated high-frequency noise is illustrated in Fig.~\ref{fig:modillus}(d) after artificially modulating the carrier signal.

\begin{figure*}[htb!] 
\vspace{0pt}
\centering
\includegraphics[width = 0.999\textwidth]{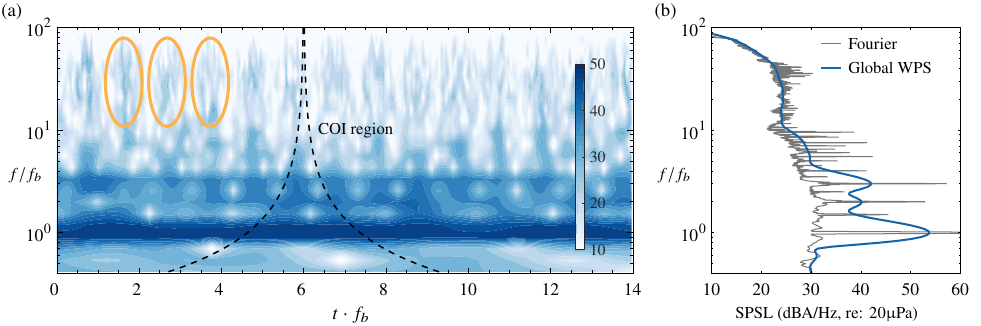}
\caption{(a) Wavelet power spectrum (WPS) of a rotor noise time series, corresponding to a case with a relatively high degree of BPFM (microphone C, indicated in Fig.~\ref{fig:expAC}(b), for an advance ratio of $J = 0.61$). (b) Global, time-averaged WPS associated with the same data used to construct sub-figure (a), in comparison to a conventional Fourier-based spectrum.}
\label{fig:timefreq1}
\end{figure*}

It is important to realize that BPFM does not affect an ensemble-averaged spectrum (or any other second-order statistics). However, the intensity variations are audible, even when the BPF is inaudible to the human ear (this is the case especially for large-scale rotorcraft systems for which typically $f_b < 20$\,Hz). Preserving the temporal dimension of the data can be done explicitly by performing a time-frequency analysis. By way of a wavelet transform using a Morlet wavelet, a wavelet power-spectrum (WPS) of a noise signal can be generated. Such a time-frequency representation of the acoustic energy is shown in Fig.~\ref{fig:timefreq1}(a) for a time series that is subject to BPFM (details of the implementation can be found here \cite{baars:2014a2}). Alongside, in Fig.~\ref{fig:timefreq1}(b), the ensemble-averaged Fourier spectrum and time-averaged WPS are shown for this stationary noise signal. Note that the Morlet wavelet has a relatively high temporal resolution---at the expense of a fine spectral resolution---with a small cone-of-influence (COI) region. With sufficient resolution in time, it is evident that the noise at $f > 10f_b$ exhibits a strong intensity modulation with a time-scale equal to that of the blade passages (indicated by the alternating attenuation and intensification of energy). An in-depth route to separate tonal and broadband components is also useful to explore (particularly when harmonics appear at very high frequencies). In this regard, cyclo-stationary spectral analysis \cite{jurdic:2009a} or wavelet-based methods \cite{meloni:2023a} can be considered, but these are beyond the scope of the current paper.

\subsection{Present contribution and outline}\label{sec:contr}
\noindent In this study, the time- and/or ensemble-averaged characterization of rotor noise is augmented by utilizing metrics that preserve the temporal variation in the intensity of high-frequency noise \cite{baars:2021c}. This intensity-variation is a nonlinear frequency-interaction and refers, in the context of our current work, to the phase-consistencies between the low-frequency BPF and higher-frequency noise. A primary outcome of the study by Baars \emph{et al.} \cite{baars:2021c} was that for a rotor in hover, BPFM is strongest at angles of $\theta \approx -20^\circ$. A low-fidelity listening experiment also indicated that the modulation metrics are well-correlated with the degree of modulation that is audible. In terms of source mechanisms, the working principle of BPFM may initially be thought of as a result of the harmonic variation in the source-receiver distance (as a direct result of the rotor spinning). But, if that were to be the case, the normalized modulation strength would be maximum at sideline angles; this is not observed in measurements \cite{baars:2021c}. As such, a proper quantification of BPFM and an identification of its source mechanisms is needed. It will inspire new measurement and post-processing procedures to support noise regulations and the assessment of the noise impact on communities \cite{rizzi:2020tr,greenwood:2022a}. For instance, this can be achieved by including BPFM aspects in realistic auralizations of rotorcraft noise \cite{krishnamurthy:2019c,krishnamurthy:2021c} and future noise prediction tools \cite{bian:2019c,han:2020a,roger:2020a,thurman:2023c}.

This paper starts by describing an aeroacoustic experiment of a laboratory-scale rotor at advance ratios ranging from $J = 0$ to $0.61$ (\S\,\ref{sec:exp}). Then, \S\,\ref{sec:mod} covers a brief review of quantifying BPFM, after which directivity patterns of BPFM are described in tandem with the results of the rotor's inflow disturbances. Since certain noise sources become more (or less) dominant with variations in $J$ \cite{grande:2022a}, due to the changing separated flow-features around the blade at relatively low Reynolds-numbers, these data allow for identifying the source mechanisms responsible for BPFM, which are described in \S\,\ref{sec:source}.

\section{Experimental data of a small-scale rotor}\label{sec:exp}
\subsection{Experimental setup and rotor operating conditions}\label{sec:rotor}
\noindent Two experimental campaigns were conducted: (1) acoustic measurements in the rotor's near- and far-field regions, and (2) flow-field measurements using Particle Image Velocimetry (PIV).\footnote{All acoustic data are archived open-access and are available online at: \url{https://doi.org/10.4121/9c8cf649-7617-42e2-a9b1-a32d5f483964}. Flow-field data are available upon request: please email the authors at \url{w.j.baars@tudelft.nl}.} All aeroacoustic measurements were conducted in the anechoic A-Tunnel \cite{merino:2020a} of the Delft University of Technology. This facility is anechoic at frequencies above 200\,Hz. Internal dimensions are roughly 6.4\,m (L) $\times$ 6.4\,m (W) $\times$ 3.2\,m (H). The wind tunnel inlet measures 0.6\,m in diameter and provides a uniform, low-turbulence intensity inflow velocity ($\sqrt{\overline{u^2}}/U_\infty \lesssim 0.05$\,\%). Atmospheric pressure, temperature and relative humidity were practically constant throughout the measurement duration and were taken as $p_\infty = 101\,325$\,Pa, $T_\infty = 293.15$\,K and $\rm{RH} = 40$\,\%, respectively, yielding a density of $\rho_\infty = 1.207$\,kg/m$^3$ and a sound speed of $a_\infty = 343.2$\,m/s.

\begin{figure*}[htb!] 
\vspace{0pt}
\centering
\includegraphics[width = 0.999\textwidth]{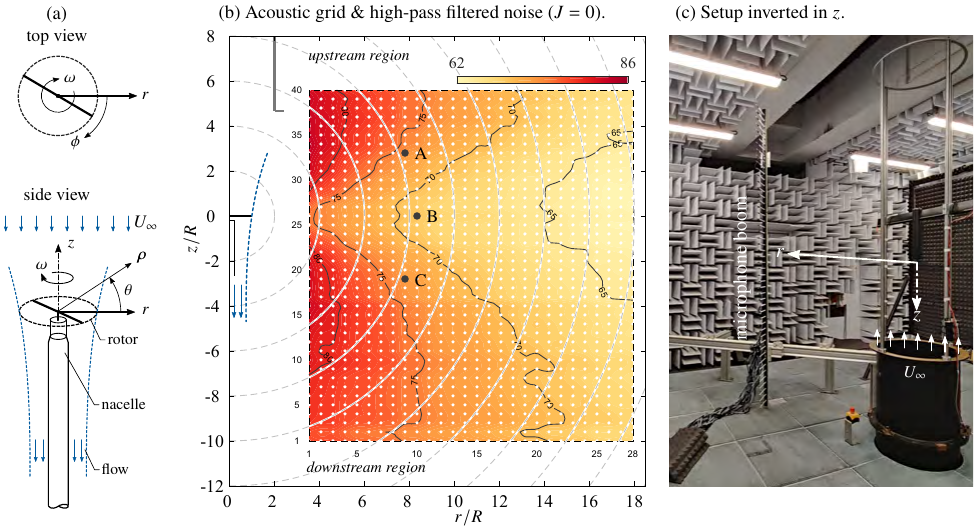}
\caption{(a,c) Setup with the rotor axis $z$ and radial coordinate $r$; the rotor hub is at $(r,z) = (0,0)$. (b) Acoustic grid of 1\,120 microphone positions mapped out with a linear microphone boom ($m = 1:40$ microphone positions, and $b = 1:28$ boom positions). The acoustic contour is described in the text. Microphones A, B and C are used later and are situated at $\rho \approx 4.2D_p$ and $\theta \approx 19.7^\circ$, $\theta = 0^\circ$ and $\theta \approx -19.7^\circ$, respectively.}
\label{fig:expAC}
\end{figure*}

A rotor test rig was mounted to the circular wind tunnel inlet, supporting a small-scale rotor in hover (see Fig.~\ref{fig:expAC}c). A circular nacelle of 5\,cm in diameter embedded a compact 6-axis ATI Mini40 sensor (with maximum thrust and torque capacities of 40\,N and 1\,Nm, respectively), providing rotor thrust and torque readings. An LMT 2280 brushless motor was used in combination with a TDK-Lambda power supply to drive the rotor, comprising a voltage range of 0--60\,V and a current range of 0--80\,A. A US Digital EM1 transmissive optical encoder, coupled with a US Digital disk of 25\,mm in diameter, gave a one-per-revolution (1P) signal of the rotor shaft for an accurate reading of its rotational speed and angular position. The induced flow direction was physically upward in the facility, but in the plots throughout the paper the orientation is flipped upside down to represent a rotor-in-hover scenario when $J = 0$. Finally, any flow recirculation was not observed qualitatively in this relatively large chamber with a large anechoically-treated exhaust slit located roughly $6D_p$ downstream of the rotor (Fig.~\ref{fig:expAC}c); it is therefore expected that an intensification of BPF harmonics---that have previously been linked to a flow recirculation in anechoic chambers \cite{stephenson:2019a}---are absent.

The rotor itself was derived from an APC propeller (model \href{https://www.apcprop.com/product/9x6e/}{9x6e}); this rotor has a diameter of 9\,inch and a pitch of 6\,inch. For the current setup, the diameter was scaled up to $D_p = 2R = 0.30$\,m, while all blade elements were re-shaped with a NACA 4412 airfoil. The rotor, made of an aluminum alloy, was manufactured in-house using CNC machining and with a 0.4 to 0.8\,$\upmu$m Ra finish. This rotor is identical to the one used in benchmarking studies (\emph{i.e.}, BANC\,X) focusing on the flow transition over the blades and its influence on the aeroacoustic performance \cite{casalino:2021a,grande:2022a,grande:2022ba}. The rotor spun at a nominal rate of $\omega = 131.0$\,rev/s (7\,860\,RPM), resulting in a BPF of $f_b = 262.0$\,Hz. Rotor-rotational speed was kept constant to within $\pm 0.1$\% with the aid of a closed-loop PID-type controller working with the 1P signal. For the hover condition, the Reynolds number was $Re_{c75} \equiv c_{75} 2\pi\omega 0.75R/\nu = 1.35 \cdot 10^5$, based on the blade chord of $c_{75} = 22.4$\,mm at $r = 0.75R$, and the tip Mach number was $M_{\rm tip} \equiv 2\pi\omega R/a_\infty = 0.358$ (note that the rotational speed of the rotor in rad/s is denoted as $\Omega = 2\pi\omega$). Predicting the noise can be difficult at low Reynolds-numbers \cite{deters:2014c,casalino:2021a,gojon:2021a,casalino:2022a}, when the rotor operates with a reduced efficiency. Hence, the experimental data can aid the validation of simulations.

To consider a variation in rotor-blade loading (as a representation of varying forward flight velocity with fixed RPM rotors), various advance ratios were considered. The advance ratio is defined as $J \equiv U_\infty/(\omega 2R) = \pi U_\infty/U_{\rm tip}$, and four different values were considered by fixing the rotational speed of the rotor, while changing the tunnel inflow (freestream) velocity, $U_\infty$. For the four advance ratios considered, being $J = 0$ (hover), 0.24, 0.41 and 0.61, the corresponding freestream velocities were $U_\infty \approx 0$, 9.6, 16 and 24\,m/s, respectively. Tip Mach numbers and Reynolds numbers were nearly constant (listed in Table~\ref{tab:expcond}) and considered a total velocity composed of the rotor's rotational speed component and the incoming freestream velocity component (\emph{e.g.,} $M_{\rm tip} = U_{\rm tip}/a_\infty$, with $U_{\rm tip} = \sqrt{(\Omega R)^2 + U_\infty^2}$). Measurements of the thrust force ($F_z$) and rotor torque ($\tau_z$) were performed for each advance ratio to yield rotor-performance data (Table~\ref{tab:expcond}). Coefficients of thrust ($C_T$) and torque ($C_\tau$), and the propulsive efficiency ($\eta_p$), were calculated following the relations,
\begin{eqnarray}
 \label{eq:ctcq}
 C_T = \frac{F_z}{\rho A \Omega^2 R^2},~~~~C_\tau = \frac{\tau_z}{\rho A \Omega^2 R^3},~~~~\eta_p = \frac{JC_T}{\pi C_\tau}.
\end{eqnarray}
Here $A = \pi R^2$ is the rotor disk area; rotor power in Watts is taken as $P = \Omega\tau_z$. Finally, for the hover condition only, the rotor's figure of merit (FM) was computed following the conventional definition,
\begin{eqnarray}
 \label{eq:FM}
 {\rm FM} = \frac{C_T^{\nicefrac{3}{2}}}{\sqrt{2} C_\tau}.
\end{eqnarray}

\begin{figure*}[htb!] 
\vspace{0pt}
\centering
\includegraphics[width = 0.999\textwidth]{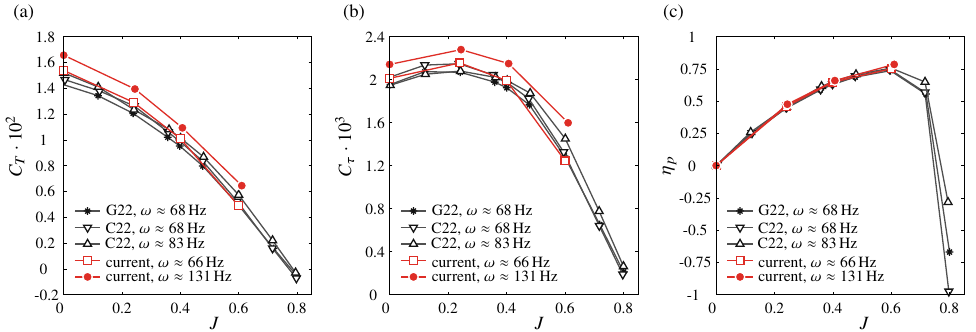}
\caption{Rotor operating performance data. (a) Thrust coefficient, $C_T$, (b) torque coefficient, $C_\tau$, and (c) propulsive efficiency, $\eta_p$, as a function of the advance ratio, $J$. Current performance data are compared to studies of the same rotor and within the same facility, but at lower rotational frequencies (G22: Grande \emph{et al.} \cite{grande:2022a} and C22: Casalino \emph{et al.} \cite{casalino:2022a}).}
\label{fig:rotorperf}
\end{figure*}

Thrust and torque coefficients, as well as the propulsive efficiency, are compared to the literature in Fig.~\ref{fig:rotorperf}. Data from the literature (black symbols) considered the same rotor and facility, except for lower rotational frequencies. As such, our current measurements also considered a lower rotational frequency for a direct comparison ($\omega \approx 66$\,Hz, with the open red markers) and these match well with the previous data. At the higher rotational speed considered in this paper (solid red markers), the rotor operates at a higher thrust coefficient as the consequence of a more turbulent (slightly less separated) flow; this comes at the expense of a larger torque coefficient although the propulsive efficiency remains equal.

At hover, the thrust and torque coefficients, as well as the FM, can be compared to a parametric study on small-scale rotors by Tinney \& Sirohi \cite{tinney:2018a}. Even though the FM is equal, our thrust and torque coefficients are higher by about 30\,\% and 40\,\%, respectively. This is presumably caused by the more aggressive rotor pitch (roughly 9\,inch in our study, compared to 4.5\,inch in the study by Tinney \& Sirohi \cite{tinney:2018a}). Deviations in thrust (3\,\%) and torque (33\,\%), from the manufacturer's performance \href{https://www.apcprop.com/files/PER3_11x10E.dat}{data}, are attributed to simplifications in the theoretical predictions used to generate those performance data.

When considering non-zero advance ratios, the thrust coefficient is known to decrease with an increase in $J$, due to the lower rotor disk loading. The torque coefficient stays roughly constant for $J < 0.4$ due to the separated flow close to the root of the blades, resulting in large drag values as reported by Grande \emph{et al.} \cite{grande:2022a}. The propulsive efficiency does increase with $J$ \cite{brandt:2011c,grande:2022a}, since the blade section angle of attack reduces below the stall angle and the torque decreases sufficiently fast so that the rotor efficiency peaks at $J \approx 0.6$ (these studied considered $\omega \approx 67$\,rev/s, but the performance trends are thus similar for our current rotational speed of $\omega = 131.0$\,rev/s).
\begin{table}[htb!] 
 \begin{center}
  \caption{Operating conditions of the $D_p = 0.30$\,m diameter rotor, for each of the four advance ratios, $J$ (the rotor is spinning at the same nominal rate of 7\,860\,RPM, while the freestream velocity $U_\infty$ is varied to change $J$).}
  \label{tab:expcond}
  \resizebox{\linewidth}{!}{%
  \begin{tabular}{c|cccc|ccccccc}
  $J$ & $U_\infty$ (m/s) &  $f_b$ (Hz) & $M_{\rm tip}$ & $Re_{c75}$ & $F_z$ (N) & $\tau_z$ (Nm) & $P$ (W) & $C_T \cdot 10^{2}$ & $C_\tau \cdot 10^{3}$ & $\eta_p$ & FM\\
  \hline
  0 & 0 & 262.0         & 0.358 & $1.36 \cdot 10^5$ & 21.3 & 0.41 & 339 & 1.66 & 2.14 & 0 & 0.70\\
  0.24 & 9.6 & 262.0    & 0.359 & $1.36 \cdot 10^5$ & 17.9 & 0.44 & 361 & 1.39 & 2.28 & 0.48 & -- \\
  0.41 & 16.0 & 262.0   & 0.361 & $1.38 \cdot 10^5$ & 14.0 & 0.41 & 341 & 1.09 & 2.15 & 0.66 & -- \\
  0.61 & 24.0 & 262.0   & 0.365 & $1.40 \cdot 10^5$ & 8.3 & 0.41 & 253 & 0.65 & 1.60 & 0.78 & -- \\
  \end{tabular}}
 \end{center}
\end{table}

\subsection{Measurements of the acoustic field and rotor-induced flow field}\label{sec:meas}
Acoustic data were acquired using a linear microphone boom with 40 sensors, comprising an equidistant spacing of 60\,mm. The vertical boom was mounted to a horizontal beam so that it could be traversed in $r$. Free-field microphones were oriented such that their measuring diaphragms were co-planar with the measurement plane (this orientation avoids having to point the normal vector of the diaphragm to an aeroacoustic sound source location that can be ambiguous \cite{viswanathan:2006a,fievet:2016a}). A free-field microphone correction was applied in all spectral analyses to account for the intrusive nature and form factor of the microphone ($90^\circ$ grazing incidence waves), although it only marginally affects the intensity at $f > 10$\,kHz. For each advance ratio $J$, the acoustic field was mapped out by translating the microphone boom to 28 radial positions (traversing step of 80\,mm), resulting in a total of $28 \times 40 = 1\,120$ positions for which pressure time series $p(r,z;t)$ are available (see Fig.~\ref{fig:expAC}b). 

The sensors used were $\sfrac{1}{4}$\,in. free-field microphones (\href{https://www.grasacoustics.com/products/special-microphone/array-microphones/product/830-gras-40ph-10-ccp-free-field-array-microphone}{GRAS 40PH}), with a frequency response range of 5\,Hz to 20\,kHz with a $\pm 2$\,dB accuracy (and a $\pm 1$\,dB accuracy from 50\,Hz up to 5\,kHz) and with a dynamic range of 32\,dBA to 135\,dB, with a sensitivity of 50\,mV/Pa. Microphones were calibrated in situ with a GRAS 42AA piston-phone. All 40 microphones were IEPE powered and simultaneously sampled with several NI PXIe-4499 sound and vibration modules (on-board filtering prior to digitization with a 24-bit accuracy). All signals were sampled at a rate of $f_s = 51.2$\,kHz for a duration of $T = 40$\,seconds ($2T\omega \approx 10\,480$ blade passages); this was confirmed to be more than sufficient for converged bispectral statistics at the lowest frequency of interest (see \cite{poloskei:2018a} and Appendix~A). For spectral-based analysis, the one-sided spectrum is taken as $\phi_{pp}(r,z;f) = 2\langle P(r,z;f) P^*(r,z;f)\rangle$, where $P(r,z;f) = \mathcal{F}\left[p(r,z;t)\right]$ is the temporal FFT and $\langle\cdot\rangle$ denotes ensemble-averaging. Sound pressure spectrum levels (SPSL) in dB/Hz follow ${\rm SPSL}(r,z;f) = 10\log_{10}(\phi_{pp}(r,z;f)/p_{\rm ref}^2)$, with $p_{\rm ref} = 20\,\upmu$Pa. Ensemble-averaging was conducted using FFT partitions of $N = 16f_s/\omega$ samples, to ensure that the discrete frequencies align with the BPF and its harmonics; this reduces the leakage of tonal energies into neighbouring frequencies \cite{tinney:2018a}. The value of $N$ yields a spectral resolution of ${\rm d}f = 8.2$\,Hz and 653 ensembles with 50\,\% overlap.
\begin{figure*}[htb!] 
\vspace{0pt}
\centering
\includegraphics[width = 0.999\textwidth]{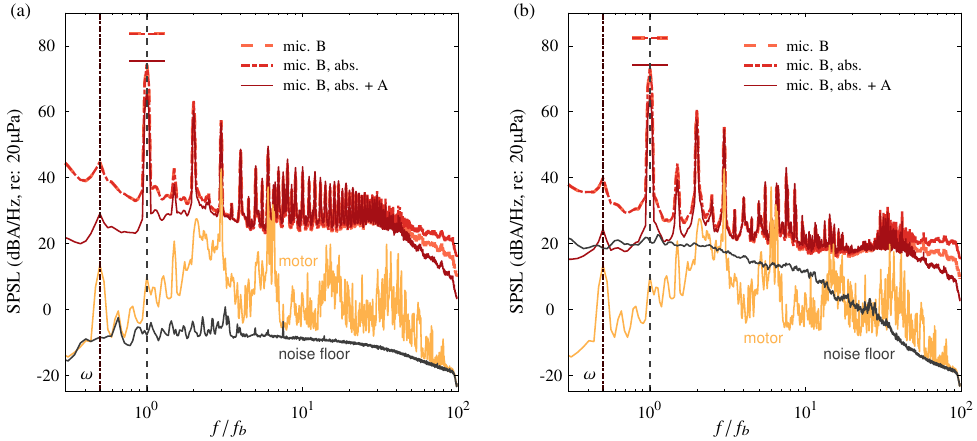}
\caption{Acoustic spectra for (a) $J = 0$, and (b) $J = 0.41$, corresponding to microphone B (indicated in Fig.~\ref{fig:expAC}b). Plots include spectra of the raw acoustic time series (dashed), a spectrum after correction for atmospheric absorption (dash-dotted), and a spectrum with an additional A-weighting applied (solid). SPSL magnitudes of the BPF are indicated with horizontal bars and correspond to the amplitude of a pure tone at $f = f_b$ (see text). Spectra corresponding to the motor-only scenario and the noise floor of the facility (and wind tunnel noise in the case of $J \neq 0$ is) are also shown.}
\label{fig:spectracor}
\end{figure*}

For all analyses, the raw pressure time series were subject to a band-pass filter, with a flat response between 60\,Hz and 15\,kHz, suppressing the non-anechoic, low-frequency content and the energy beyond the highest frequency range of the microphone. The SPSL was also corrected for atmospheric absorption (ANSI S1.26-1996) with an assumed propagation distance from the rotor hub (primarily affecting $f > 10$\,kHz). In addition, an A-weighting was applied (ANSI S1.6-1967) to account for the relative loudness perceived by the human ear \cite{fletcher:1933a} and attenuates energy at $f \lesssim 3f_b$. Finally, it was confirmed that motor-only noise was far less intense than the rotor noise. As an example, the acoustic pressure spectra shown in Figs.~\ref{fig:spectracor}(a,b) correspond to microphone B, for $J = 0$ and 0.41, respectively. Raw and corrected pressure spectra of the rotor noise are shown, as well as the corrected noise spectra of only the motor (spinning without rotor, but at the same RPM) and the noise floor of the facility (with only the wind tunnel running for the $J \neq 0$ cases). The motor noise alone has the expected spectral peak at $f = \omega = f_b/2$ and is also relatively dominant in the vicinity of $f = 14\omega = 7f_b$, which is caused by the 14 magnetic poles of the motor (note that the motor noise is also known to arise from structural vibrations and harmonic interference \cite{huff:2018c,henderson:2018c}). Nevertheless, the magnitude of the noise floor and motor-only noise are low compared to the rotor noise.

\begin{figure*}[htb!] 
\vspace{0pt}
\centering
\includegraphics[width = 0.999\textwidth]{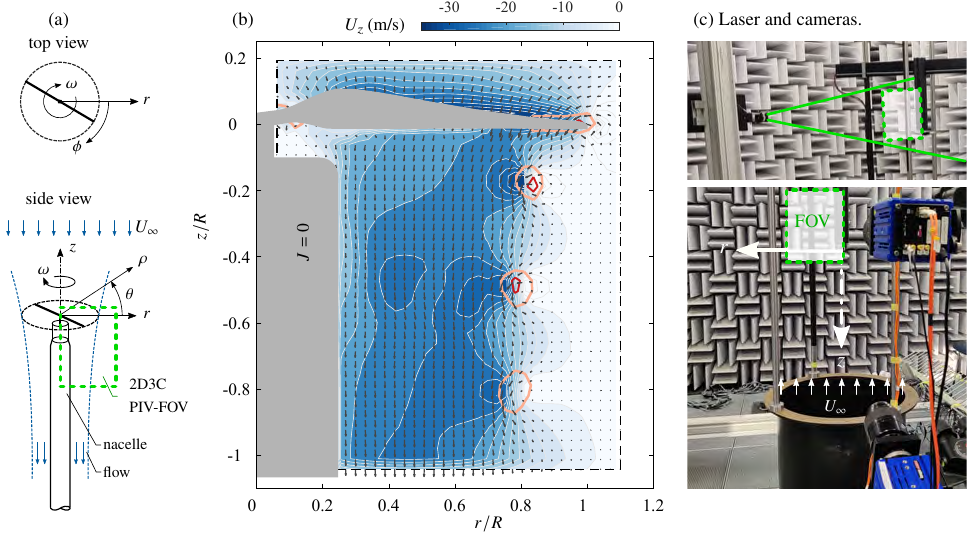}
\caption{(a,c) Setup with the PIV field-of-view. (b) Contour of the phase-averaged vertical velocity $U_z$ at a near-zero phase angle behind the blade, for $J = 0$; vectors show the in-plane velocity with a vector-skip of 15. Superimposed are two red iso-contours of in-plane vorticity at magnitudes of $\omega_\phi = 1\,500$\,s$^{-1}$ and $7\,500$\,s$^{-1}$.}
\label{fig:expPIV}
\end{figure*}

Flow fields were captured using a stereoscopic PIV setup (Fig.~\ref{fig:expPIV}) that is similar to the one reported by Grande \emph{et al.} \cite{grande:2022a,grande:2022ba}. With a double cavity Quantel Evergreen EVG00200 Nd:YAG laser, a sheet was created to illuminate a relatively large region on one side of the rotor axis. Our current PIV field-of-view (FOV) aids in detailing the vortical flow structures in the near-vicinity of the rotor blade. Two Imager sCMOS cameras were used with $2560 \times 2160$\,px$^2$. Two Nikon lenses were mounted with Scheimpflug adapters, each with a 60\,mm focal length and an f\# of 11. In this work, we only consider 500 image-pairs that were acquired in a phase-locked sense using the 1P signal. A sample result of the PIV campaign is shown in Fig.~\ref{fig:expPIV}(b) and is described in the caption.

\subsection{Sound pressure level statistics}\label{sec:splstat}
\begin{figure*}[htb!] 
\vspace{0pt}
\centering
\includegraphics[width = 0.999\textwidth]{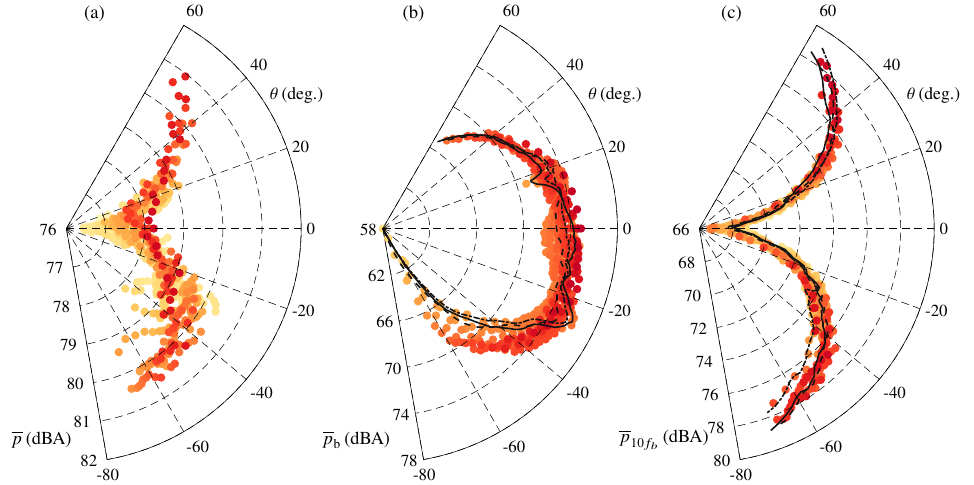}
\caption{Noise directivity patterns for advance ratio $J = 0$, obtained via spherically projecting all 1\,120 data points to a rotor-hub-centered arc of radius $\rho = 5D_p$. Sub-figures correspond to the (a) OASPL, (b) SPSL of the BPF, and (c) integrated SPSL over $f > 10f_b$ (points are colored per the unprojected noise levels).}
\label{fig:arcJ0}
\end{figure*}

Basic features of the acoustic data are documented here to generate an appreciation of the different noise characteristics of the rotor, with varying advance ratio $J$. For brevity, the case of $J = 0$ is chosen to present a few intermediate results. First, recall that a spatial topography of the integrated SPSL over $f > 10f_b$ was shown in Fig.~\ref{fig:expAC}(b). This integrated, high-pass filtered noise content is referred to as $\overline{p}_{10f_b}$ and its spatial contour clearly illustrates that the higher-frequency broadband noise exhibits the well-known quadrupole directivity pattern \cite{marte:1970tr,brentner:2003a}. Throughout the remainder of the paper the $f > 10f_b$ range is used in defining the high-frequency noise content. Changes in the $10f_b$ threshold do not affect the conclusions made. The acoustic data can be condensed to directivity arcs for ease of interpretation: SPSL statistics corresponding to all 1\,120 measurement points were projected to a rotor-hub-centered\footnote{Since the noise is, on-average, emitted from the centre location of the spinning rotor-blades, this hub-centered source position resulted in the best collapse of the data in Figs.~\ref{fig:arcJ0}(b,c).} arc with a virtual radius of $\rho = 5D_p$. In this projection, a spherical spreading law was adopted, according to $p \propto 1/\rho$. Thus, a data collapse implies that all measurement points were located in the acoustic far-field. Results for $\overline{p}_{10f_b}$ are presented in Fig.~\ref{fig:arcJ0}(c), showing an excellent collapse of the data. This is further accentuated with the set of three lines (solid, dash-dotted, dashed), resulting from a fit to the projected data from the three most outward upper-vertical-lower perimeters of locations spanned by the microphone grid, respectively. Variations are less than $\approx 1$\,dBA and these variations are more pronounced for large angles $\theta$, for which the microphones on the vertical boom were relatively close to the walls of the anechoic chamber. Withal, collapse of $\overline{p}_{10f_b}$ data is expected: the acoustic wavelength corresponding to $10f_b$ is $\lambda/D_p = a_\infty/(10f_b)/D_p \approx 0.43$, meaning that all microphone locations were situated in the acoustic far-field, except for the closest measurement points at a region around $\theta = 0^\circ$ (this is the angle at which the collapse of the data markers in Fig.~\ref{fig:arcJ0}(c) is least good).

When performing the same data-projection procedure for the OASPL (integral of the spectra over the entire frequency range), we obtain the directivity shown in Fig.~\ref{fig:arcJ0}(a). The absence of collapse suggests that some locations, in terms of the OASPL, were not yet situated in the acoustic far-field. When considering the SPSLs of only the BPF content\footnote{Here, the SPSL magnitude is taken as the root-mean-square amplitude of a time series resembling the \emph{pure} BPF tone as was shown in Fig.~\ref{fig:modillus}(c), being $A/\sqrt{2}$ for a harmonic wave with amplitude $A$. By computing the SPSL magnitude this way, instead of taking the peak of the spectrum at $f = f_b$, it is ensured that the magnitude is \emph{independent} of the chosen spectral resolution ${\rm d}f$. In practice, the amplitude is obtained by taking the magnitude of ${\rm SPL}(r,z;f) = 10\log_{10}(\phi_{pp}(r,z;f){\rm d}f/p_{\rm ref}^2)$ at $f = f_b$. Note that $\phi_{pp}$ in dBA/Hz is premultiplied by ${\rm d}f$, to obtain the BPF peak amplitude in dBA.}, and performing again a spherical projection to a virtual arc with a radius $\rho = 5D_p$, we obtain the SPSL pressure denoted as $\overline{p}_b$ (Fig.~\ref{fig:arcJ0}b). Rotor-thickness noise, as well as the loading noise from thrust and torque, classifies primarily as an acoustic dipole source, and its radiated noise is strongly confined to a region around the rotor plane. For the thrust-producing rotor this periodic noise has a maximum intensity that is oriented slightly towards the downstream direction. Since the BPF has an acoustic wavelength of $\lambda/D_p = a_\infty/f_b/D_p \approx 4.3$, only a subset of data points lie within the acoustic far-field. Data points closer to the rotor are subject to evanescent pressure waves from the source.

\begin{figure*}[htb!] 
\vspace{0pt}
\centering
\includegraphics[width = 0.999\textwidth]{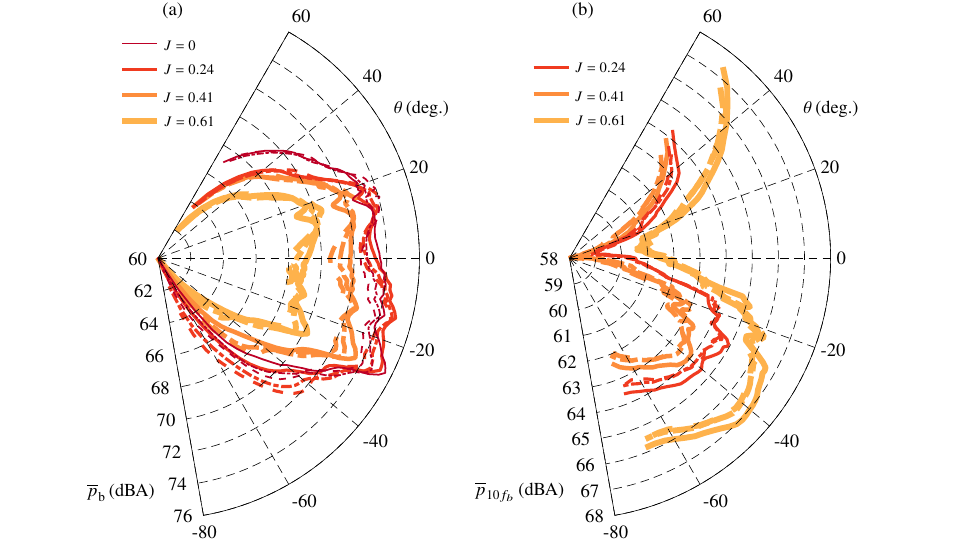}
\caption{Noise directivity patterns, as per Figs.~\ref{fig:arcJ0}(b,c), but now for all four advance ratios $J$. Each set of three lines (solid, dash-dotted, dashed) resembles a fit to the projected data from the three most outward upper-vertical-lower perimeters of the microphone grid. For sub-figure (b) with the integrated SPSL over $f > 10f_b$ the directivity pattern of $J = 0$ is omitted to keep the radial scale condensed (the $J = 0$ case was already shown in Fig.~\ref{fig:arcJ0}c).}
\label{fig:arcJs}
\end{figure*}

Noise directivity patterns of $\overline{p}_b$ and $\overline{p}_{10f_b}$ are now generated for all four advance ratios $J$, following the same procedure as described above for $J = 0$. In order to visualize the result, only the curves found through a fitting procedure to the projected data from the most far-field locations are considered. Again, these fits are done to the three most outward upper-vertical-lower perimeters of locations spanned by the microphone grid. Results are shown in Figs.~\ref{fig:arcJs}(a,b) for the BPF content and high-frequency noise, respectively. Two primary trends are observed. Firstly, the BPF tone reduces in amplitude due to the lower intensity of the thickness/loading noise (an increase in advance ratio causes a direct decrease in disk loading). Secondly, the high-frequency noise reduces in overall amplitude too, but does not exhibit a monotonic decrease with an increase in $J$. That is, the high-frequency noise content decreases for $J = 0 \rightarrow 0.24 \rightarrow 0.41$, but then increases again for the highest advance ratio of $J = 0.61$. This complex behaviour is related to noise sources associated with the change in separated-flow features over the blade \cite{grande:2022a,casalino:2022a}. With increasing $J$, the separation goes from a fully laminar separation ($J = 0$), to one that re-attaches and forms a laminar separation bubble ($J = 0.24$ and $0.41$), to one that fully separates in a turbulent state ($J = 0.61$). For the latter case, the (trailing-edge) noise of the shedding is more intense than for the laminar separation, resulting in an increase in noise intensity compared to the two intermediate advance ratios. The reason why the high-frequency noise is so dominant for the $J = 0$ case (recall the magnitude of the directivity pattern shown in Fig.~\ref{fig:arcJ0}c) is because of the turbulence-ingestion noise through the onset of a weak blade-vortex interaction (in which the blades encounter an imprint of the tip-vortex deployed by the consecutive blade).

Finally, this section merely illustrates the quality of the data and the fact that acoustic data were taken in both the acoustic near- and far-field. Note however that the BPFM analysis of \S\,\ref{sec:mod} is unaffected by the pressure obeying (or not obeying) a far-field spreading trend and its amplitude decay rate. Specifically, the modulation metrics are correlation-based and thus energy-normalized.

\subsection{Flow field in the blade's near-vicinity}\label{sec:flowfield}
\begin{figure*}[htb!] 
\vspace{0pt}
\centering
\includegraphics[width = 0.999\textwidth]{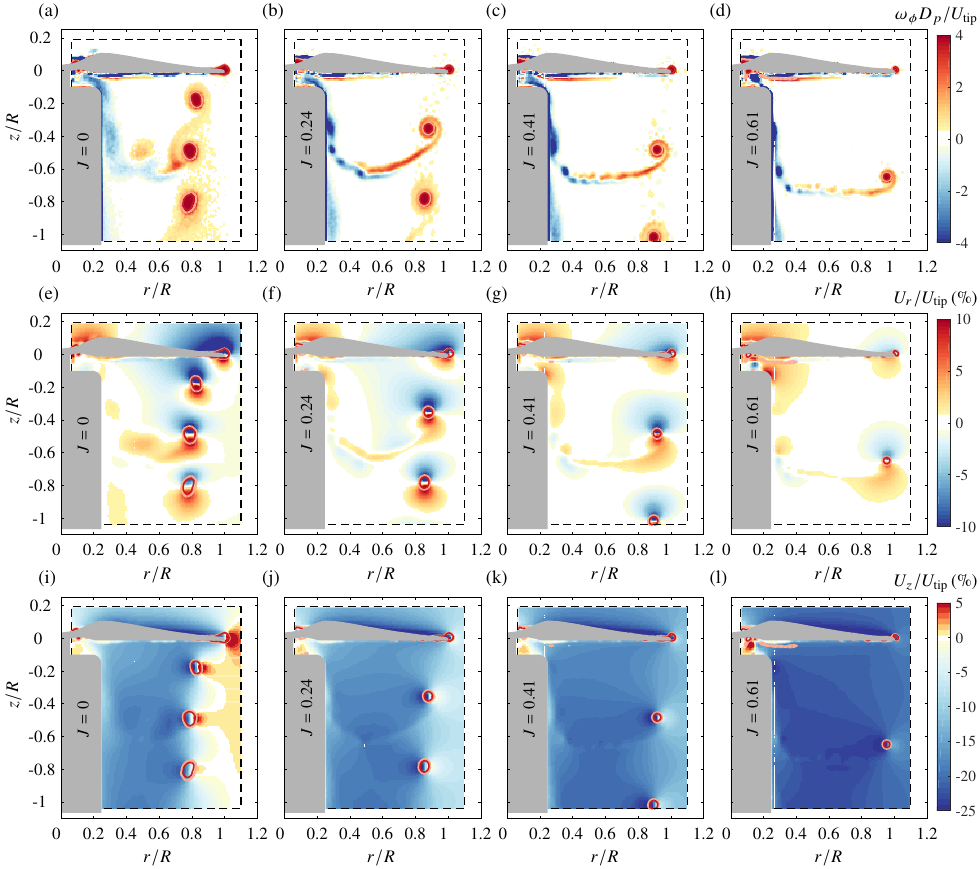}
\caption{Filled contours of the (a-d) in-plane vorticity $\omega_\phi$, (e-h) horizontal velocity $U_r$, and (i-l) vertical velocity $U_z$, for all four advance ratios $J$. Superimposed on all plots are two red iso-contours of in-plane vorticity, $\omega_\phi$, corresponding to normalized vorticity magnitudes of $\omega_\phi D_p/U_{\rm tip} =  3.5$ and $7.0$.}
\label{fig:PIVJsbig}
\end{figure*}

Results of PIV are shortly described here to provide an overview of the primary flow features induced by (and encountered by) the rotor blade. Phase-locked fields of various flow quantities at a near-zero phase angle behind the blade are shown in Fig.~\ref{fig:PIVJsbig}, for all four advance ratios $J$. Fields of the in-plane vorticity $\omega_\phi$ are shown in Figs.~\ref{fig:PIVJsbig}(a-d). Locations of the spiralling tip vortices are well-identified, as well as the vorticity in the wake sheet of the blade. For the three non-zero advance ratios, this blade wake connects the tip vortex at a wake age of $180^\circ$ to the rotor hub location where a relatively weak root vortex `folds' around the nacelle. For the $J = 0$ case, the tip vortices and wake sheets have a small axial spacing, causing an interaction of the wake sheets associated with the $180^\circ$ and $360^\circ$ wake ages (as also described in detail in the work by Thurman \emph{et al.} \cite{thurman:2023a}). As expected for lower blade loading, the vorticity magnitude decreases with an increasing value of $J$.  Figs.~\ref{fig:PIVJsbig}(e-h) and Figs.~\ref{fig:PIVJsbig}(i-l) show filled contours of the horizontal velocity $U_r$ (positive outboard) and vertical velocity $U_z$ (negative downward). Superimposed on these two sets of plots are two iso-contours of the in-plane vorticity, $\omega_\phi$, to highlight the location of the tip vortices. It becomes apparent that the vortical flow field leading to the onset of the tip-vortex causes an inboard motion of the flow on the blade's suction side and outboard on the blade's pressure side. For $J = 0$, the inboard horizontal velocity reaches more than 10\,\% of the rotor-tip velocity. This high velocity magnitude is the consequence of the contracting slipstream and an inboard flow induced by the preceding blade's tip vortex situated with its core less than $0.2R$ below the rotor plane. For this advance ratio, turbulence-ingestion noise will be dominant. Even for the $J = 0.24$ case the tip vortex with a $180^\circ$ wake age trails relatively close to the pressure side of the blade and an imprint of this vortical flow still appears clearly at the bottom edge of the blade near $r/R \approx 0.6$. Only for $J \geq 0.41$ the influence of the vortical flow from the preceding blade does no longer affect the flow around the successive blade, in terms of a horizontal velocity disturbance in the phase-locked mean-sense. Finally, see Grande \emph{et al.} \cite{grande:2022a} for detailed PIV-based velocity fields in the cross-section of the blades at $r/R = 0.6$.

\section{Blade passing frequency modulation}\label{sec:mod}
\subsection{Phase-averaged acoustic signature}\label{sec:phaseavg}
\begin{figure*}[htb!] 
\vspace{0pt}
\centering
\includegraphics[width = 0.999\textwidth]{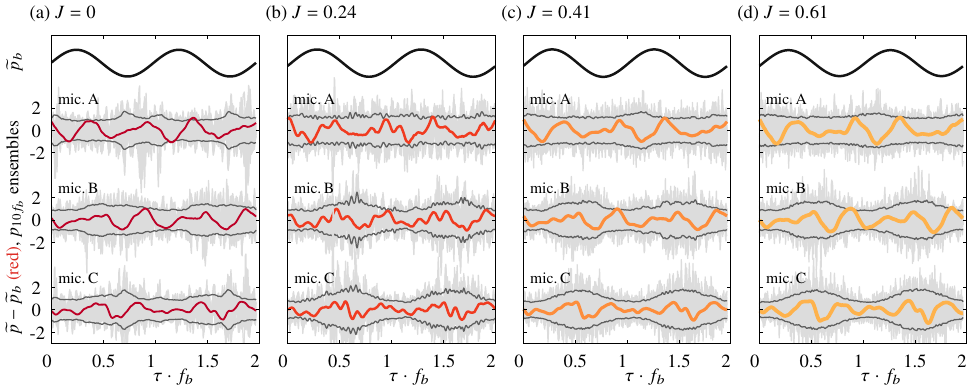}
\caption{Phase-averaged acoustic pressure signatures. For all four advance ratios $J$, from top-to-bottom: BPF tone $\widetilde{p}_b$; phase-averaged pressure signal in the absence of the BPF tone, $\widetilde{p} - \widetilde{p}_b$, in red; phase-averaged envelope of the high-pass filtered signal $p_{10f_b}$ (dark grey) and 15 ensembles superimposed (light grey). All graphs are for one full revolution of the two-bladed rotor.}
\label{fig:phasetime}
\end{figure*}

Here we make a first attempt to visualize BPFM using phase-averaged pressure data, denoted as $\widetilde{p}(r,z;\tau)$ with $\tau$ being the time coordinate within one full rotor-revolution. Signatures of the phase-averaged pressure are shown in Fig.~\ref{fig:phasetime}. First we show for each advance ratio $J$ the phase-averaged pressure time series for (1) the BPF tone $\widetilde{p}_b$, and (2) the phase-averaged pressure in the absence of the BPF (thus $\widetilde{p}-\widetilde{p}_b$, with the red thick line). Since microphones A, B and C are located at the same radial distance $\rho$ (Fig.~\ref{fig:expAC}b), the time series affiliated with the BPF are nearly identical at all three locations. Signature $\widetilde{p}-\widetilde{p}_b$ is the resultant of the 2$^{\rm nd}$ and higher-order harmonics being phase-consistent with the BPF, thus surviving the phase-average. The finer undulations within the signal (for instance clear in the mic.~B signal in Fig.~\ref{fig:phasetime}b) correspond to roughly the 7$^{\rm th}$ harmonic (postulated to be motor noise as discussed earlier). By construction, the signature does not contain any (phase-inconsistent) broadband noise. Nevertheless, BPFM can still be visualized in an easy manner: in the background of the red curves are 15 ensembles of the raw acoustic pressure, high-pass filtered at $f > 10f_b$. In addition, upper and lower envelopes to the intensity of these raw time series are shown, which were created using a Hilbert transform and by considering all 5\,240 rotation-ensembles. A strong link between the variation in the high-frequency noise intensity and the BPF signal is clearly noticeable. For the mic.~C signal (particularly for $J = 0.24$, 0.41 and 0.61), the intensity variation is well-correlated to the BPF signal, while this correlation is less strong for mic.~A.

\begin{figure*}[htb!] 
\vspace{0pt}
\centering
\includegraphics[width = 0.999\textwidth]{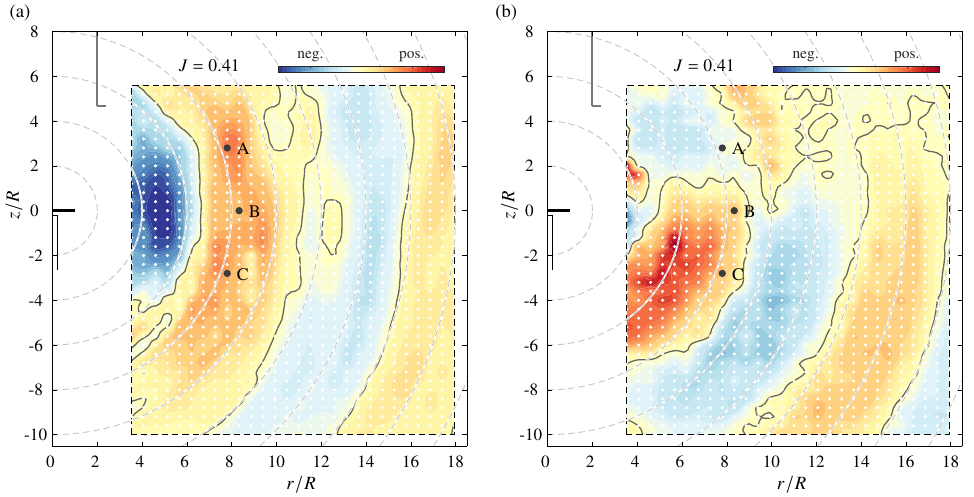}
\caption{For the case of $J = 0.41$, spatial topography of: (a) the phase-averaged pressure signal $\widetilde{p}(r,z;\tau)$ for one specific value of $\tau$, and (b) the phase-averaged zero-mean envelope of the high-pass filtered signal $p_{10f_b}$. Temporal evolution of these spatial iso-contours of acoustic pressure for one full rotor revolution, $\tau \in [0,2f_b)$, are available as movies in the supplementary material.}
\label{fig:PAcontour}
\end{figure*}

To visualize BPFM throughout the spatial extent captured by the acoustic data, similar analyses as the ones performed to construct Fig.~\ref{fig:phasetime} can be performed for each microphone. The phase-averaged pressure data allows for visualizing a spatial contour of $\widetilde{p}(r,z;\tau)$ for one specific value of $\tau$ (note that movies with temporal evolution, $\tau \in [0,2f_b)$, are available as supplementary material). Fig.~\ref{fig:PAcontour}(a) shows this contour for the case of $J = 0.41$. Evidently, the BPF is dominant with a spatial wavelength of around $4.3D_p$; finer undulations superimposed on the BPF are also visible, as well as the decaying nature of the sound wave amplitude. Alongside, in Fig.~\ref{fig:PAcontour}(b), a contour is shown of the envelope-amplitude to the intensity of the high-pass filtered signal at $f > 10f_b$ (thus equal to the zero-mean envelopes shown in Fig.~\ref{fig:phasetime}). Certainly below the rotor disk plane, a strong link exists between the BPF signal and the variation in the high-frequency noise. Trends follow Fig.~\ref{fig:phasetime}(c) in that the link appears strongest along a radial path going through microphone C, while being weakest for a path going through microphone A. Moreover, the phase shift between the BPF and the envelope changes drastically when moving from locations above the rotor disk plane to ones below it. Next, BPFM will be quantified with correlation-based scalar metrics.

\subsection{Scalar metrics and directivity patterns of modulation}\label{sec:patterns}
Baars \emph{et al.} \cite{baars:2021c} described methods for quantifying BPFM. The various scalar metrics involved in this process are briefly summarized below, and it must be emphasized that these are computed from a single acoustic pressure time series. Then, the metrics are computed for every acoustic pressure time series of the grid measurements to inspect the trends of the modulation in the acoustic near- and far-field regions.\\[-10pt]

\begin{enumerate}[label=(\roman*),itemsep=0.5pt,topsep=1pt,leftmargin=0.75cm]
\item \noindent Through a bispectral analysis, the dominant quadratic inter-frequency coupling can be found out of all possible frequency combinations present within a signal. This analysis effectively correlates two frequency components, $f_1$ and $f_2,$ to their sum ($f_3 = f_1 + f_2$) or difference and can be expressed as an auto-bicoherence, $\gamma^2_{ppp}(f_1,f_2)$, according to:
\begin{eqnarray}\label{eq:bicoh1}  
    \gamma^2_{ppp} = \frac{\vert \phi_{ppp}\left(f_1,f_2\right) \vert^2}{\phi_{pp}\left(f_1\right)\phi_{pp}\left(f_2\right)\phi_{pp}\left(f_1+f_2\right)} \in [0,1].
\end{eqnarray}
Here, the numerator is the cross-bispectrum, taken as $\phi_{ppp}\left(f_1,f_2\right) = 2\langle P\left(f_1+f_2\right) P^*\left(f_1\right) P^*\left(f_2\right)\rangle$. Coordinates $r$ and $z$ are omitted for ease of notation. Note that $\gamma^2_{ppp}(f_1,f_2)$ indicates the degree of normalized correlation between the energy at $f_1$ and $f_2$, and the energy at $f_1 + f_2$ (here we only consider sum-interactions, and not the difference-interactions per $f_3 = f_1 - f_2$, as we are interested in how the low-frequency BPF modulates higher-frequency noise). A sample auto-bicoherence spectrum is shown in Appendix~A, corresponding to the time series of microphone C at conditions of $J = 0$ (Fig.~\ref{fig:appbispecJ1}) and $J = 0.41$ (Fig.~\ref{fig:appbispecJ3}). Typically, a ridge of relatively strong correlation appears along $f_2 = f_b$, meaning that the BPF is phase-coupled to a broad range of frequencies within the same signal, $f_1 > f_b$. Said quadratic coupling is suppressed in phase-averaging (\S\,\ref{sec:phaseavg}) because the \emph{phase} in the cross-bispectrum can still vary per triad. A single metric $\Gamma^2_{m}$ is constructed by averaging $\gamma^2_{ppp}(f_1,f_2)$ for the primary frequency $f_2 = f_b$ and all possible quadratic frequency doublets residing at $f_1 > 10f_b$ (see Appendix~A for the full details).\\[-10pt]
\item \noindent The concept of \emph{modulating-} and \emph{carrier}-signals allows for the application of standard linear correlation methods. The modulating signal $p_b(t)$ is taken as the BPF-associated time series, while a carrier signal $p_h(t)$ is taken as the time series resulting from high-pass filtering at $f > 10f_b$. An envelope capturing the time-varying intensity of the latter signal can be generated through a Hilbert transform $\widehat{p}_h(t) = \vert H[p_h(t)]\vert$. By correlating the modulating signal $p_b(t)$ with the carrier envelope $\widehat{p}_h(t)$, we obtain the temporal cross-correlation $R_a(\tau_c) = \langle p_b(t) \widehat{p}_h(t-\tau_c) \rangle$; when normalised with the standard deviations we obtain the normalised correlation coefficient, $\rho_a(\tau_c) \in [0,1]$. Finally, two modulation metrics are defined: the correlation strength, $\rho_a = {\rm max}[\rho(\tau_c)]$, and the phase $\phi_a = \tau_{cm} f_b(2\pi)$, where $\tau_{cm}$ is the temporal shift for which the maximum correlation value occurs.\\[-10pt]
\end{enumerate}

\begin{figure*}[htb!] 
\vspace{0pt}
\centering
\includegraphics[width = 0.999\textwidth]{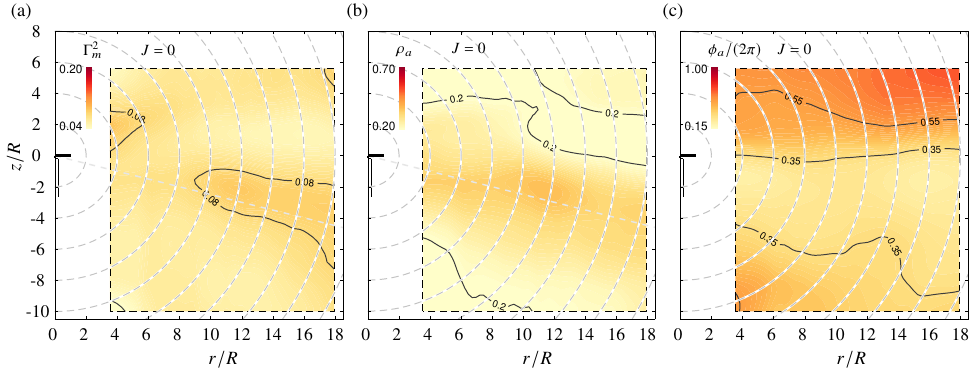}
\caption{For the case of $J = 0$, spatial topographies of: (a) the auto-bicoherence-based metric $\Gamma^2_{m}$, (b) the modulation strength $\rho_a$, and (c) the relative phase $\phi_a$ between modulating \& carrier signals. A reference (dashed) line corresponding to $\theta = -14^\circ$ is indicated in sub-figures (a) and (b).}
\label{fig:gridmodJ1}
\end{figure*}

\begin{figure*}[htb!] 
\vspace{0pt}
\centering
\includegraphics[width = 0.999\textwidth]{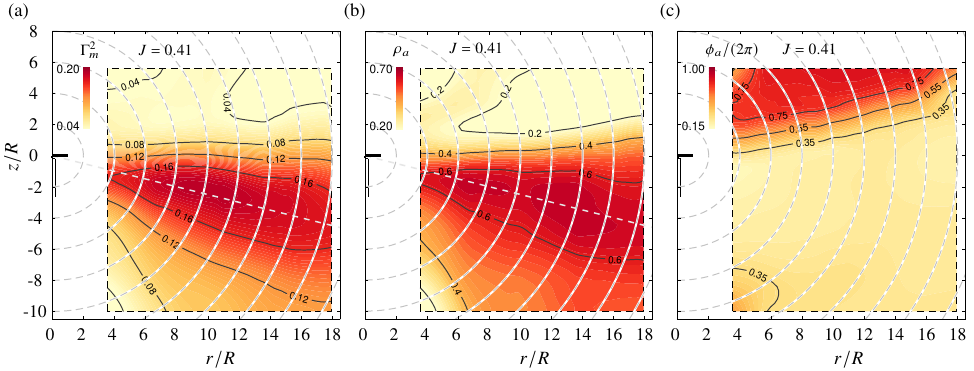}
\caption{Similar to Fig.~\ref{fig:gridmodJ1}, but now for $J = 0.41$.}
\label{fig:gridmodJ3}
\end{figure*}

\noindent Note that all modulation metrics considered in the current work are correlation-based and energy-normalized ($\Gamma^2_{m}$, $\rho_a$ and $\phi_a$). Hence the acoustic pressure-decay is irrelevant, and the analysis is applicable to a single acoustic pressure time series anywhere in the acoustic near- or far-field regions. All three metrics are computed for each of the 1\,120 acoustic signals and the results for two advance ratios, $J = 0$ and 0.41, are shown in Figs.~\ref{fig:gridmodJ1} and~\ref{fig:gridmodJ3}, respectively.

It is apparent that the BPFM strength, captured by metrics $\Gamma^2_{m}$ and $\rho_a$, show a similar pattern for $J = 0.41$, thus highlighting the robustness of the two metrics. Moreover, the strength is considerably weaker for $J = 0$. Later on, in this section, we detail the difference in BPFM strength for changes in the advance ratio. When we here focus on the $J = 0.41$ case with its pronounced BPFM strength, it is seen that the metrics are maximum for $\theta \approx -14^\circ$ (grey dashed line in Figs.~\ref{fig:gridmodJ3}a,b). This will result in a distinguishable \emph{``wop-wop"} or \emph{``buzzing"} character of the noise at that polar angle. The strength of BPFM remains constant with outward distance. Even though this is expected for a pure convective acoustic wave field, it was never shown explicitly. It furthermore validates the quality of the acoustic data in the free-field simulated environment: \emph{e.g.}, the modulation is \emph{not} related to nodes or anti-nodes caused by reflections and acoustic waves interfering in an in- and/or out-of-phase manner. 

The phase relationship between modulating and carrier signals is shown in Fig.~\ref{fig:gridmodJ3}(c) and is important for auralization methodologies \cite{krishnamurthy:2019c}. An out-of-phase behaviour corresponds to a phase of $\phi_a = 0.5(2\pi)$, which would mean that the BPF signal leads to the occurrence of highest-intensity in $p_{10f_b}$ by half-a-period of the BPF. For the sector where the BPFM strength is large (say $-45^\circ \lesssim \theta \lesssim 0^\circ$), the phase in Fig.~\ref{fig:gridmodJ3}(c) equals roughly $\phi_a = 0.25(2\pi)$: the BPF signal thus leads the associated, periodic intensity variation in the high-frequency noise by a quarter period of the BPF (this is also apparent from the mic.~C signal in Fig.~\ref{fig:phasetime}c).

It is important to realize that $\Gamma^2_{m}$ is derived from the magnitude of the auto-bicoherence (and here does not include phase information of the underlying auto-bispectrum). As such, the strength of BPFM in the classical sense (a buzzing/breathing of higher-frequency noise at a rate of the BPF) is quantified by $\rho_a$ and $\phi_a$ in the remainder of this work. That is, the phase of the frequency content in the `total' carrier signal, relative to the modulating signal, was preserved in the two-point correlation analysis through the identification of $\phi_a$, while $\Gamma^2_{m}$ is a measure of the phase-coupling on a per-frequency basis.

\begin{figure*}[htb!] 
\vspace{0pt}
\centering
\includegraphics[width = 0.999\textwidth]{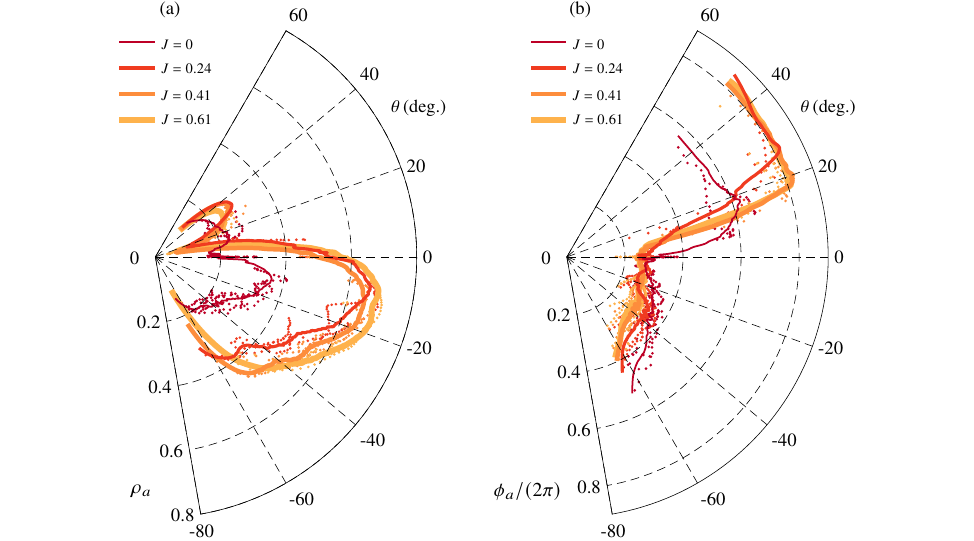}
\caption{Directivity patterns of BPFM, for all four advance ratios $J$, with in (a) the modulation strength $\rho_a$, and (b) the relative phase $\phi_a$ between modulating \& carrier signals.}
\label{fig:arcJsmod}
\end{figure*}

In order to infer the influence of the advance ratio $J$ on BPFM, similar results were constructed as the ones shown in Figs.~\ref{fig:gridmodJ1} and~\ref{fig:gridmodJ3}, but now for all four advance ratios. Since it was furthermore confirmed that the metrics were invariant with outward radial distance $\rho$, the directivity patterns of the metrics are presented with polar plots. Figs.~\ref{fig:arcJsmod}(a) and~\ref{fig:arcJsmod}(b) show the directivity patterns of $\rho_a$ and $\phi_a$, respectively. Individual data points are shown with small markers, as well as a fit line through these data for each $J$.

Observations on the BPFM-directivity patterns can be summarized as follow, and will be described in terms of the underlying source mechanisms in \S\,\ref{sec:source}.\\[-10pt]

\begin{enumerate}[label=(\roman*),itemsep=0.5pt,topsep=1pt,leftmargin=0.75cm]
\item \noindent Current directivity patterns confirm earlier observations of BPFM for a hovering rotor ($J = 0$): modulation is predominantly present within the downstream sector, with a maximum strength towards $\theta \approx -20^\circ$. Still, a small degree of BPFM is present above the rotor disk plane.
\item \noindent For the non-zero advance ratios the BPFM remains weak above the rotor. The phase reaches a near-constant value of $\phi_a = 0.8(2\pi)$ and is indicative of a slight temporal `lead' of the high-frequency noise intensity, relative to the BPF.\\[-10pt]
\item \noindent Below the rotor plane, the BPFM is significantly stronger for the non-zero advance ratios. In the sector of $-45^\circ \lesssim \theta \lesssim -5^\circ$ the BPFM strength slightly increases from $J = 0.24$ to the highest advance ratio tested, $J = 0.61$, although the overall directivity pattern is very similar. The maximum strength resides around $\theta = -14^\circ$ for all non-zero $J$, and the associated phase remains roughly constant at a temporal lag of the high-frequency intensity, of $\phi_a \approx 0.25(2\pi)$.\\[-10pt]
\end{enumerate}

\section{Source mechanisms of BPF modulation}\label{sec:source}
\begin{figure*}[htb!] 
\vspace{0pt}
\centering
\includegraphics[width = 0.999\textwidth]{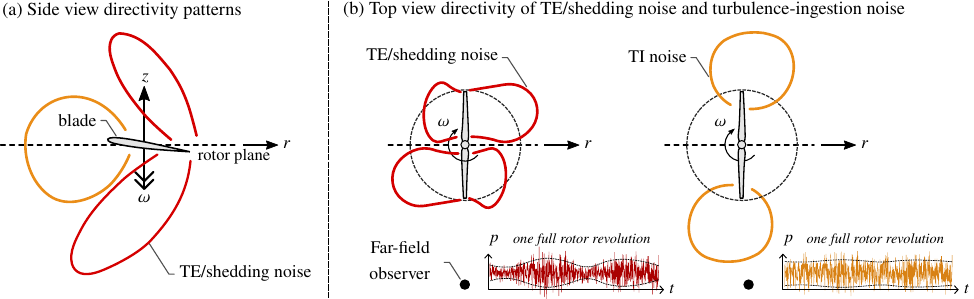}
\caption{Schematic illustration of directivity patterns in the context of high-frequency noise. Both trailing edge (TE) and flow shedding-type, as well as turbulence-ingestion (TI) noise due to the onset of blade-vortex interaction (BVI) noise source mechanisms are considered. A far-field observer is subject to a higher (relative) degree of high-frequency modulation when the TI source is absent.}
\label{fig:BPFMsum}
\end{figure*}

One driving factor involved in the BPFM can be thought of as the periodic variation in the source-receiver distance, due to the advance and retreat of the rotor blades. However, this effect alone would be more effectively felt at sideline angles than in the upstream and downstream regions. Given that the BPFM strength was identified to be maximum in the downstream region, the variation in source-receiver distance cannot be the root cause of BPFM. In an attempt to unravel the mechanisms at play in generating the trends in BPFM, we first focus on the high-frequency noise content of rotor noise signals. High-frequency noise can come from two source mechanisms: (1) noise associated with turbulence-ingestion, primarily affecting the leading-edge noise source mechanisms, and (2) noise associated with trailing-edge mechanisms and the shedding of (coherent) flow features from the separated region over the blade's suction side. Here the leading-edge noise is considered `high' in frequency since it is in relation to the much lower BPF and the blade relative velocity. Fig.~\ref{fig:BPFMsum} presents a conceptual schematic of the noise directivity patterns associated with these sources, and in the reference frame that is fixed to the rotor blades. In a side view (Fig.~\ref{fig:BPFMsum}a), the trailing edge noise signature has a noise directivity that is tilted forward, while leading-edge noise is more omnidirectional. A top view of the noise directivity patterns is drawn in Fig.~\ref{fig:BPFMsum}(b) and accentuates the following characteristics of the noise. First, the source mechanisms of trailing-edge noise is considered. Its noise generating mechanism (related to flow features shedding past the trailing-edge) is relatively coherent along the span of the blade. Directivity-wise this results (per blade) in a dominant lobe forward (although still noticeable behind the blade). As a consequence of the rotor blade spinning, the far-field observer experiences a sweep through the directivity pattern and given the directive nature of this noise source, the BPFM of this noise source alone is strong. Secondly, when concentrating on the source mechanism of leading-edge turbulence ingestion, and in particular the outboard part of the blade encountering imprints of the tip vortex from the preceding blade, the noise directivity originates from the blade tip and is relatively more omni-directional due to the relatively small nature of this noise source (not present along the entire blade span). When a far-field observer experiences a sweep through the directivity pattern due to the rotating blades, the relative strength is less than for the trailing edge-noise.

\begin{figure*}[htb!] 
\vspace{0pt}
\centering
\includegraphics[width = 0.999\textwidth]{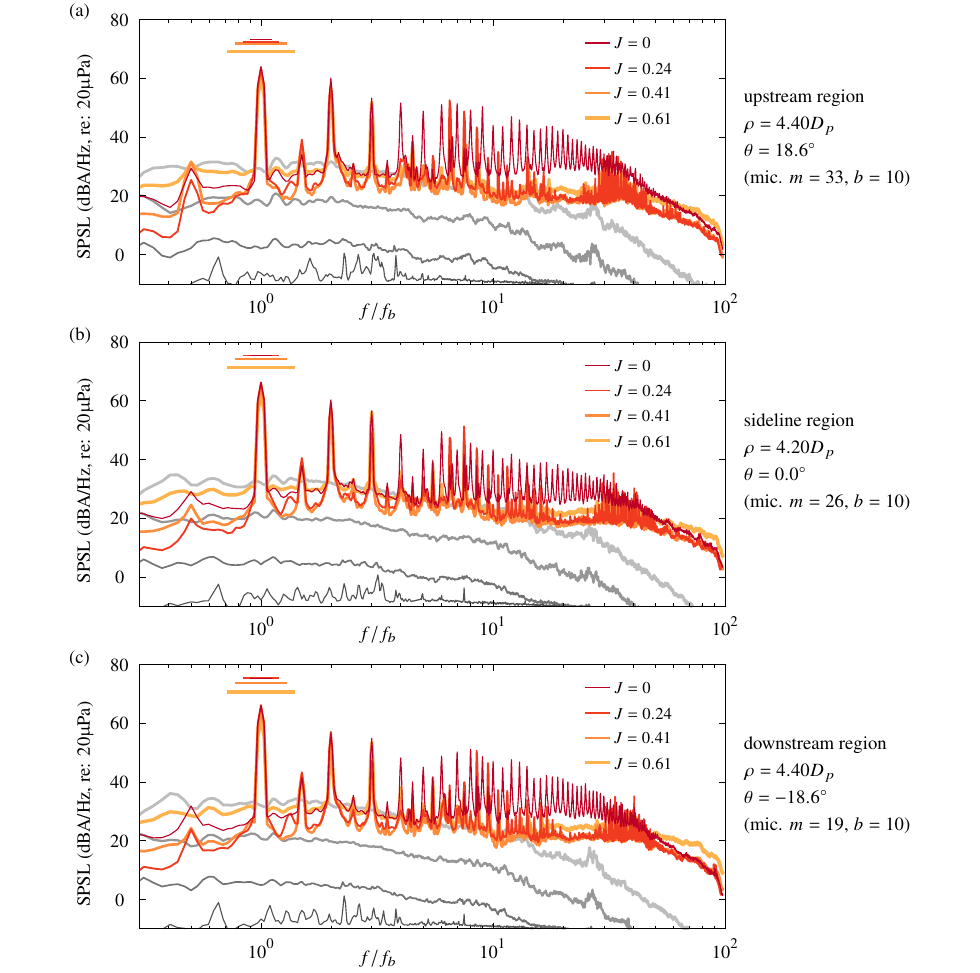}
\caption{Acoustic spectra in the (a) upstream, (b) sideline, and (c) downstream regions of the acoustic field, for all four advance ratios $J$. All spectra are corrected for atmospheric-absorption and are A-weighted. SPSL magnitudes of the BPF are indicated with horizontal bars and correspond to the amplitude of a pure tone at $f = f_b$ (see text). Spectra corresponding to the noise-floor of the facility (and wind tunnel noise in the case of $J \neq 0$ is) are also shown.}
\label{fig:spectraJ}
\end{figure*}

When we now first focus on the BPFM trends for non-zero advance ratios, the leading-edge noise source mechanism related to the onset of BVI is absent (particularly for the two highest advance ratios, recall the discussion of Fig.~\ref{fig:PIVJsbig}). As such, the modulation is strong since the high-frequency noise content is dominated by the very directive trailing-edge noise-generation mechanisms. These mechanisms include the trailing edge diffraction of the shedding of laminar separation bubbles on the blade, and/or of the flow structures in the separated region from a fully turbulent state. Specifically, the bubbles were found for the cases of $J = 0$, 0.24 and 0.41, and did increase in size when moving towards the trailing edge when the angle of attack was decreasing \cite{grande:2022a} (when the advance ratio $J$ was increasing). So for the largest separation bubbles extending all the way up to the tips of the rotor blades, the wake vortex shedding and its associated noise are dominant. For $J = 0.61$ the wake vortex shedding becomes even more dominant when there is no reattachment of the separated flow region. When inspecting the acoustic spectra this can also be observed. Fig.~\ref{fig:spectraJ} presents acoustic spectra in the upstream (Fig.~\ref{fig:spectraJ}a), sideline (Fig.~\ref{fig:spectraJ}b), and downstream regions (Fig.~\ref{fig:spectraJ}c). When focusing on the latter, the spectrum associated with $J = 0.61$ clearly has a larger amplitude than the $J = 0.24$ and 0.41 cases.

When concentrating on the hover case ($J = 0$), the directive trailing-edge noise source mechanisms must still be present. However, the noise-content at the high-frequency portion of the spectrum is dominated by the turbulence-ingestion mechanism. This can be observed from the spectra at upstream, sideline, and downstream angles. In all cases, the spectrum is dominant when considering the range $f > 10f_b$, and a multitude of harmonics (spectral-peaks) are still present. Because the high-frequency noise is dominated by this component, the perception of modulation is less due to the omni-directional noise directivity of this leading-edge mechanism, dominated by the rotor-tip blade region.

\section{Concluding remarks}\label{sec:concl}
A comprehensive study was presented on how the advance ratio (and the associated change in noise source mechanisms) influences the so-called blade passage frequency modulation (BPFM). Correlation-based metrics were successfully applied to high-fidelity acoustic datasets, spanning both the near- and far-field regions of an isolated rotor. It was hypothesized that the appearance of a strong BPFM is related to an acoustic observer---at a fixed position relative to the rotor---experiencing sweeps through the directivity pattern of trailing-edge/shedding noise (fixed to the spinning rotor blade). This type of directive noise is dominant at high-frequencies for non-zero advance ratios, and thus results in the largest degree of modulation. For the hover scenario, the high-frequency noise content is dominated by turbulence-ingestion noise, a source mechanism affiliated with the leading-edge of the rotor blade and the tip vortex from the preceding blade residing in close proximity to the successive blade. It was conjectured that the omni-directional noise directivity of this leading-edge mechanism, dominated by the rotor-tip blade region, results in a relatively weaker modulation.

Future work should strengthen our hypotheses, by way of incorporating reduced-order models of the broadband noise sources in predictive tools for the tonal noise content (\emph{e.g.}, through implementation of a compact dipole/monopole Ffowcs-Williams and Hawking's acoustic analogy \cite{fuerkaiti:2022a}, and (empirical) directivity patterns of the leading- and trailing-edge noise sources). When concerning human perception and annoyance, future work should explore the connection between the engineering BPFM metrics and psycho-acoustic metrics. Once this connection has been established, BPFM metrics can facilitate the assessment of the noise impact of AAM vehicles (by for instance applying it to data of high-fidelity numerical computations of rotor noise \cite{henricks:2020a,lee:2020a,casalino:2019c2} or other noise prediction frameworks \cite{bian:2019c,han:2020a,roger:2020a,sagaga:2020c,thurman:2023c}).

\section*{Acknowledgements}
The authors wish to gratefully acknowledge Mr. Edoardo Grande for assisting in the experiments, and for stimulating discussions about the content of this manuscript. We would also like to give special thanks to Dr.\,ir. Tomas Sinnige for setting up the power supply and the RPM-control capability of the rotor.

\section*{Appendix A: Bispectral analysis for computing a modulation metric}\label{sec:AppA}
\begin{figure*}[htb!] 
\vspace{0pt}
\centering
\includegraphics[width = 0.999\textwidth]{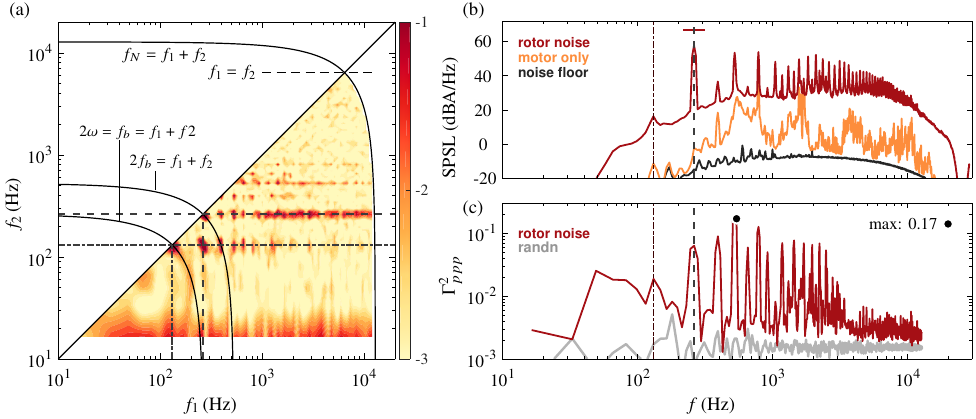}
\caption{Contour of the auto-bicoherence $\log_{10}\left[\gamma^2_{ppp}\right]$ generated from the time series of microphone C, for $J = 0$. (b) Corresponding acoustic spectra at microphone position C, with alongside in (c) the summed auto-bicoherence $\Gamma^2_{ppp}(f = f_1 + f_2)$ for the rotor noise, and for a generated reference signal comprising random noise.}
\label{fig:appbispecJ1}
\end{figure*}

Through a bispectral analysis, the dominant quadratic inter-frequency coupling can be found out of all possible frequency combinations present within a signal (here taken as $p$ and its Fourier transform, $P(f) = \mathcal{F}[p(t)]$). This analysis effectively correlates two frequency components, $f_1$ and $f_2,$ to their sum ($f_3 = f_1 + f_2$) or difference and can be expressed as an auto-bicoherence, $\gamma^2_{ppp}(f_1,f_2)$, according to:
\begin{eqnarray}\label{eq:bicoh1}  
    \gamma^2_{ppp} = \frac{\vert \phi_{ppp}\left(f_1,f_2\right) \vert^2}{\phi_{pp}\left(f_1\right)\phi_{pp}\left(f_2\right)\phi_{pp}\left(f_1+f_2\right)} \in [0,1].
\end{eqnarray}
Here, the numerator is the cross-bispectrum, taken as $\phi_{ppp}\left(f_1,f_2\right) = 2\langle P\left(f_1+f_2\right) P^*\left(f_1\right) P^*\left(f_2\right)\rangle$. Note that $\gamma^2_{ppp}(f_1,f_2)$ indicates the degree of normalized correlation between the energy at $f_1$ and $f_2$, and the energy at $f_1 + f_2$ (here we only consider sum-interactions). An sample auto-bicoherence spectrum is shown in Fig.~\ref{fig:appbispecJ1}(a) for microphone C and for $J = 0$. Typically, a ridge is shown of relatively strong bicoherence along $f_2 = f_b$. Since this means that the BPF is phase-coupled to a broad range of frequencies with the same signal (thus to frequencies $f_1 > f_b$), this ridge is representative of the degree of BPFM.

To infer what content in a time series, or auto-spectrum (plotted in Fig.~\ref{fig:appbispecJ1}b for the corresponding auto-bicoherence in Fig.~\ref{fig:appbispecJ1}a), is \emph{involved} in quadratic sum-interactions, the 2D auto-bicoherence $\gamma^2_{ppp}(f_1,f_2)$ can be condensed to a one-dimensional summed bicoherence \cite{boashash:1995bk,powers:2004m,baars:2014a3}. This is done by way of averaging along lines of constant $f = f_1 + f_2$: 
\begin{eqnarray}
 \label{eq:bicoh3}
 \Gamma^2_{ppp}\left(f\right) = \frac{1}{N_q(f)} \sum_{f = f_1 + f_2} \gamma^2_{ppp}\left(f_1,f_2\right).
\end{eqnarray}
Here $N_q(f)$ is the number of frequency doublets $f_1,f_2$. This summed bicoherence spectrum is shown in Fig.~\ref{fig:appbispecJ1}(c) and its frequency axis is aligned with the auto spectrum in Fig.~\ref{fig:appbispecJ1}(b). It is evident that $\Gamma^2_{ppp}$ shows the degree of nonlinear interactions that are buried in certain frequency components (but for their $f_1$ and $f_2$ origin it is necessary to reside back to Fig.~\ref{fig:appbispecJ1}a). On a final note, the summed bicoherence spectrum is also plotted for a generated reference signal comprising random noise (uncorrelated in a linear and nonlinear way), highlighting that the bicoherence in the rotor noise signal is significant; it was furthermore ensured that all results of our bispectral analysis are converged \cite{poloskei:2018a}.

\begin{figure*}[htb!] 
\vspace{0pt}
\centering
\includegraphics[width = 0.999\textwidth]{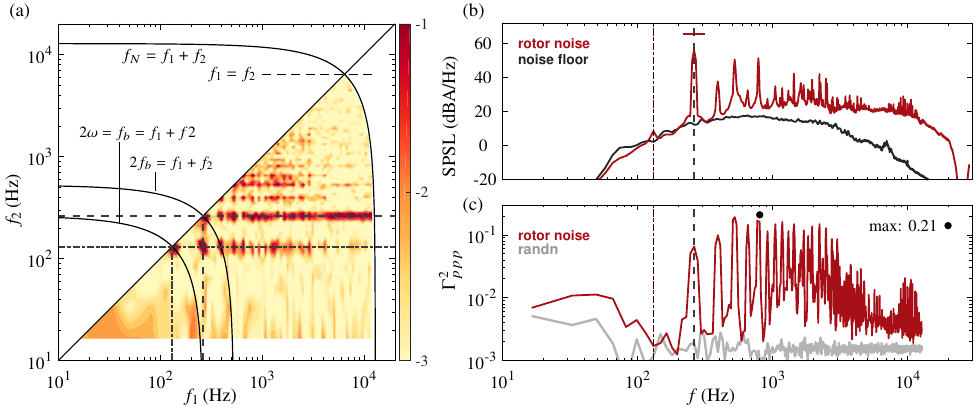}
\caption{Similar to Fig.~\ref{fig:appbispecJ3}, but now for $J = 0.41$.}
\label{fig:appbispecJ3}
\end{figure*}

For further inspection and in the context that BPFM modulation appears stronger for $J = 0.41$, in comparison to $J = 0$, plots similar to the ones presented in Fig.~\ref{fig:appbispecJ1} are shown in Fig.~\ref{fig:appbispecJ3}, but now for $J = 0.41$. It is apparent that the ridge of bicoherence along $f_2 = f_b$ is higher in magnitude, particularly for the higher-frequencies, $f_1 > 10f_b$.

\begin{figure*}[htb!] 
\vspace{0pt}
\centering
\includegraphics[width = 0.999\textwidth]{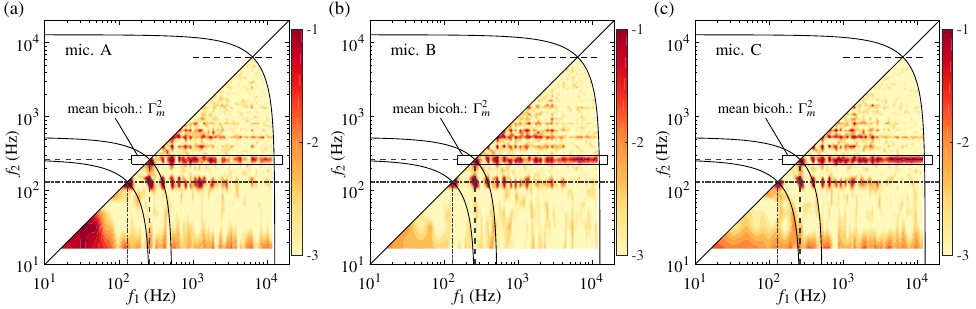}
\caption{(a,b,c) Contours of auto-bicoherence $\log_{10}\left[\gamma^2_{ppp}\right]$ at microphone positions A, B and C, for the $(f_1,f_2)$ domain comprising triadic interactions comprising the \emph{sum} of two existing interactions $f = f_1 + f_2$. The rectangular box encompassing $f_2 = f_b$ identifies the portion of the auto-bicoherence used in forming $\Gamma^2_{m}$.}
\label{fig:appbicJ3sum}
\end{figure*}

Given that the auto-bicoherence helps in forming a holistic view on the degree of phase coupling, we can define a single metric when considering the BPF tone as one of the primary frequencies forming all possible quadratic frequency doublets. For this we take the mean value of the auto-bicoherence along $f_2 = f_b$, according to
\begin{eqnarray}
 \label{eq:bicoh4}
 \Gamma^2_m  = \frac{1}{N_a} \sum_{f_1} \gamma^2_{ppp}\left(f_1,f_2 = f_b\right).
\end{eqnarray}
$\Gamma^2_m$ is a measure for the degree of phase coupling between the noise at $f > f_b$ and the BPF tone at $f = f_b$ ($N_a$ is the number of discrete points over which the auto-bicoherence is summed). From a preliminary assessment of the auto-bicoherence and parameter $\Gamma^2_m$ at positions A, B and C for $J = 0.41$ (Fig.~\ref{fig:appbicJ3sum}), it is evident that $\Gamma^2_m$ is varying in the field (\emph{e.g.}, $\Gamma^2_m$ is minimum at position A and maximum at position C).

\bibliography{Bibs/bibtex_database}
\bibstyle{aiaa}
\end{document}